\documentclass[pre,twocolumn,preprintnumbers,showpacs,showkeys,floatfix]{revtex4-1}
\usepackage{amsmath,amscd,amssymb}
\usepackage[english]{babel}
\usepackage{color}
\usepackage{graphicx}
\usepackage{comment}
\usepackage{subfigure}
\usepackage{multirow}
\usepackage[colorlinks=true, pdfstartview=FitV, 
linkcolor=blue, citecolor=blue, urlcolor=blue]{hyperref}
\usepackage{longtable}
\usepackage{lineno}

\begin{document}


\title{Modeling flow and transport in fracture networks using graphs}


\author{S.~Karra}
\email{satkarra@lanl.gov}
\affiliation{Computational Earth Science (EES-16),
Earth and Environmental Sciences Division, 
Los Alamos National Laboratory, Los Alamos, NM 87545.}
\author{D.~O'Malley}
\affiliation{Computational Earth Science (EES-16),
Earth and Environmental Sciences Division, 
Los Alamos National Laboratory, Los Alamos, NM 87545.}

\author{J.~D.~Hyman}
\affiliation{Computational Earth Science (EES-16),
Earth and Environmental Sciences Division, 
Los Alamos National Laboratory, Los Alamos, NM 87545.}

\author{H.~S.~Viswanathan}
\affiliation{Computational Earth Science (EES-16),
Earth and Environmental Sciences Division, 
Los Alamos National Laboratory, Los Alamos, NM 87545.}

\author{G.~Srinivasan}
\affiliation{Applied Mathematics and Plasma Physics (T-5), 
Theoretical Division, 
Los Alamos National Laboratory, Los Alamos, NM 87545.}


\date{\today}

\begin{abstract}
Fractures form the main pathways for flow in the subsurface within low-permeability rock. For this reason,
accurately predicting flow and transport in fractured systems is vital for
improving the performance of subsurface applications.
Fracture sizes in these systems can range from millimeters to kilometers.
Although, modeling flow and transport using the discrete fracture network
(DFN) approach is known to be more accurate due to incorporation of the
detailed fracture network structure over continuum-based methods,
capturing the flow and transport in such a wide range of scales is
still computationally intractable. Furthermore, if one has to quantify
uncertainty, hundreds of realizations of these DFN models have to be run.
To reduce the computational burden,
we solve flow and transport on a graph representation of 
a DFN. 
We study the accuracy of the graph approach by comparing breakthrough times
and tracer particle statistical data between
the graph-based and the high-fidelity DFN approaches, for fracture networks
with varying number of fractures and
degree of heterogeneity. Due to our recent developments in capabilities
to perform DFN high-fidelity simulations on fracture networks with large
number of fractures, we are in a unique position to perform such a comparison.
We show that the graph approach shows a consistent bias with up to an order of magnitude slower breakthrough when compared to the DFN approach. 
We show that this is due to graph algorithm's under-prediction of the pressure gradients across intersections on a given fracture, leading to slower tracer particle speeds between intersections and longer travel times. 
We present a bias correction methodology to the graph algorithm that reduces the discrepancy between the DFN and graph predictions. 
We show that with this bias correction, the graph algorithm predictions significantly improve and the results are very accurate. 
The good accuracy and the low computational cost, with $\mathcal{O}(10^4)$ times lower times than the DFN, makes the graph algorithm, an ideal technique to incorporate in uncertainty quantification methods.

\end{abstract}

\pacs{47.56.+r, 91.55.Jk, 91.60.Ba, 05.60.Cd, 02.10.Ox, 07.05.Tp}
\keywords{subsurface, flow, transport, graph theory, 
fractures, breakthrough, particle tracking}

\maketitle



%

%
\section{Introduction}
\label{Sec:S1_Intro}

\begin{figure}
  \begin{center}
    \includegraphics[width=0.475\textwidth]
      {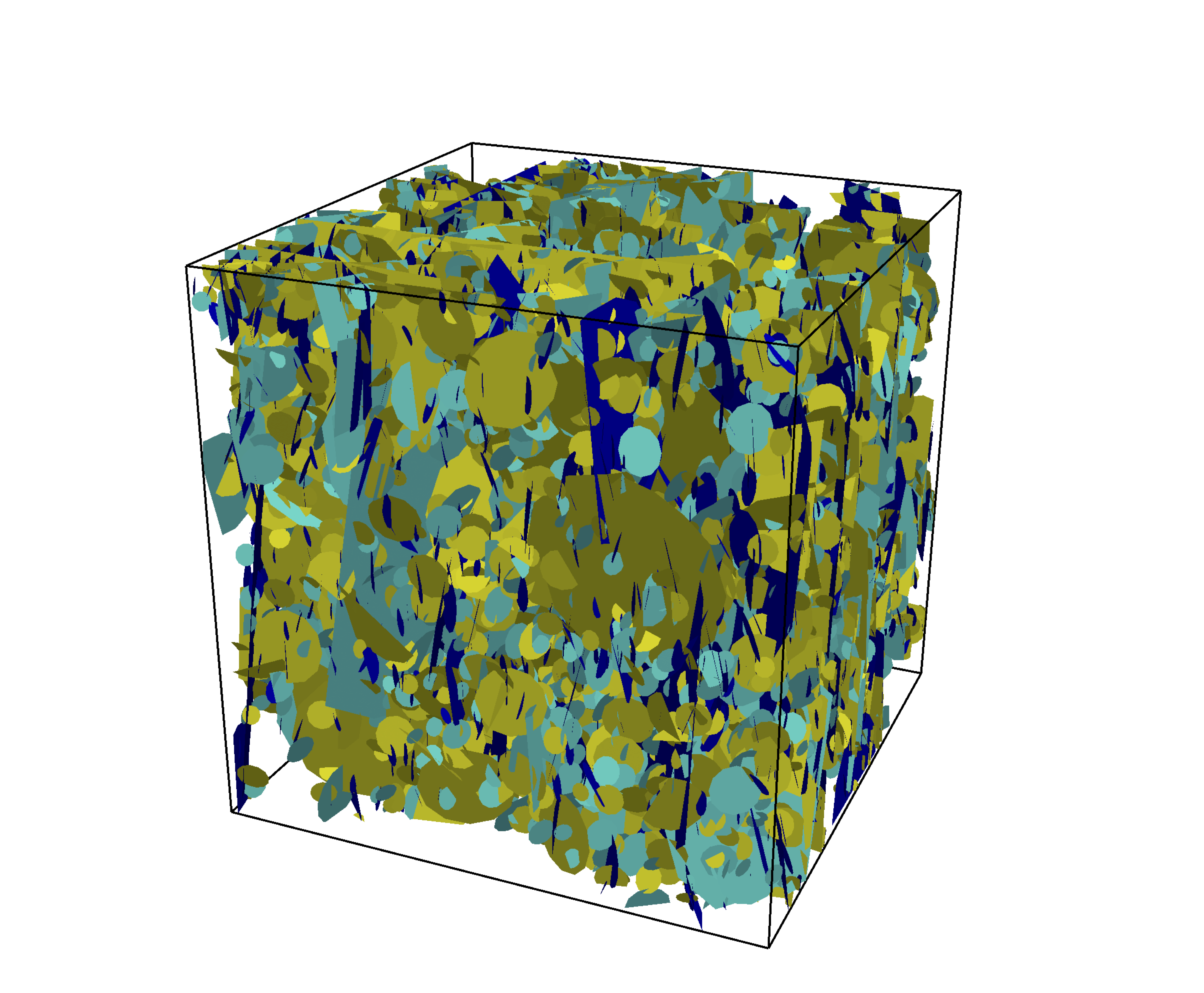}
  \end{center}
  \caption{Discrete 
  fracture network made up of 6330 circular fracture whose radii are 
  sampled from three independent truncated power-law distributions. 
  Fractures are colored by family.
  There are about 13 million grid cells in this model.} 
  \label{Fig:dfn_example}
\end{figure}

Fracture networks are the main pathways for 
fluid flow and transport in the subsurface within low-permeability rock 
\citep{adler2012fractured,national1996rock,berkowitz2002characterizing}.
Prediction of fluid migration in these fractures is
critical for several energy and national 
security applications such as hydrocarbon extraction from unconventional 
resources, geothermal energy extraction, nuclear waste disposal, and detection of 
underground nuclear explosions \citep{karra2015effect,selroos2012effect,
horne1983dispersion,jordan2014uncertainty}. The pathways formed in the
fracture networks and the fine-scale heterogeneity that they give rise
to depend heavily on the connectivity and geometrical features such 
as size and aperture of the fractures. Higher fracture density leads to
better connectivity which in turn increases
the chances for more flow and transport. Furthermore,
the larger the fracture size, the chances for connectivity
with other fractures is higher, and the larger the aperture, the more fluid volume 
can move in that fracture. Modeling
approaches have to ensure that these connectivity and geometrical features of 
fracture networks are reasonably captured for accurate predictions.
Discrete fracture network (DFN)
modeling is one such approach. In this method, fractures are represented
as two-dimensional planar objects in three-dimensional space 
(for example, see Fig.~\ref{Fig:dfn_example}), and 
flow is solved using a Darcy solver \citep{bear2013dynamics}
while transport is solved using 
an advection-dispersion equation (ADE) solver \citep{cacas1990modeling,endo1984model}
or via particle tracking \citep{schwartz1983stochastic}. The DFN method allows 
for explicit incorporation of fracture characteristics such as 
fracture size, aperture, etc., from a geological site
and one does not have to use upscaling techniques
or averaged parameters needed in continuum methods 
\citep{painter2005upscaling}. In addition, upscaling in continuum methods 
leads to tensorial parameters in the governing equations, e.g., 
tensor permeability for flow and tensor diffusivity for ADE.
One then has to seek higher order discretization techniques 
\cite{lipnikov2010monotone} to solve these governing equations,
in addition to the special care needed to handle some of the 
resulting artifacts the solution such as oscillations
\citep{cirpka1999numerical,chang2017large}.

\begin{figure*}[]
    {\includegraphics[width=0.475\textwidth]
      {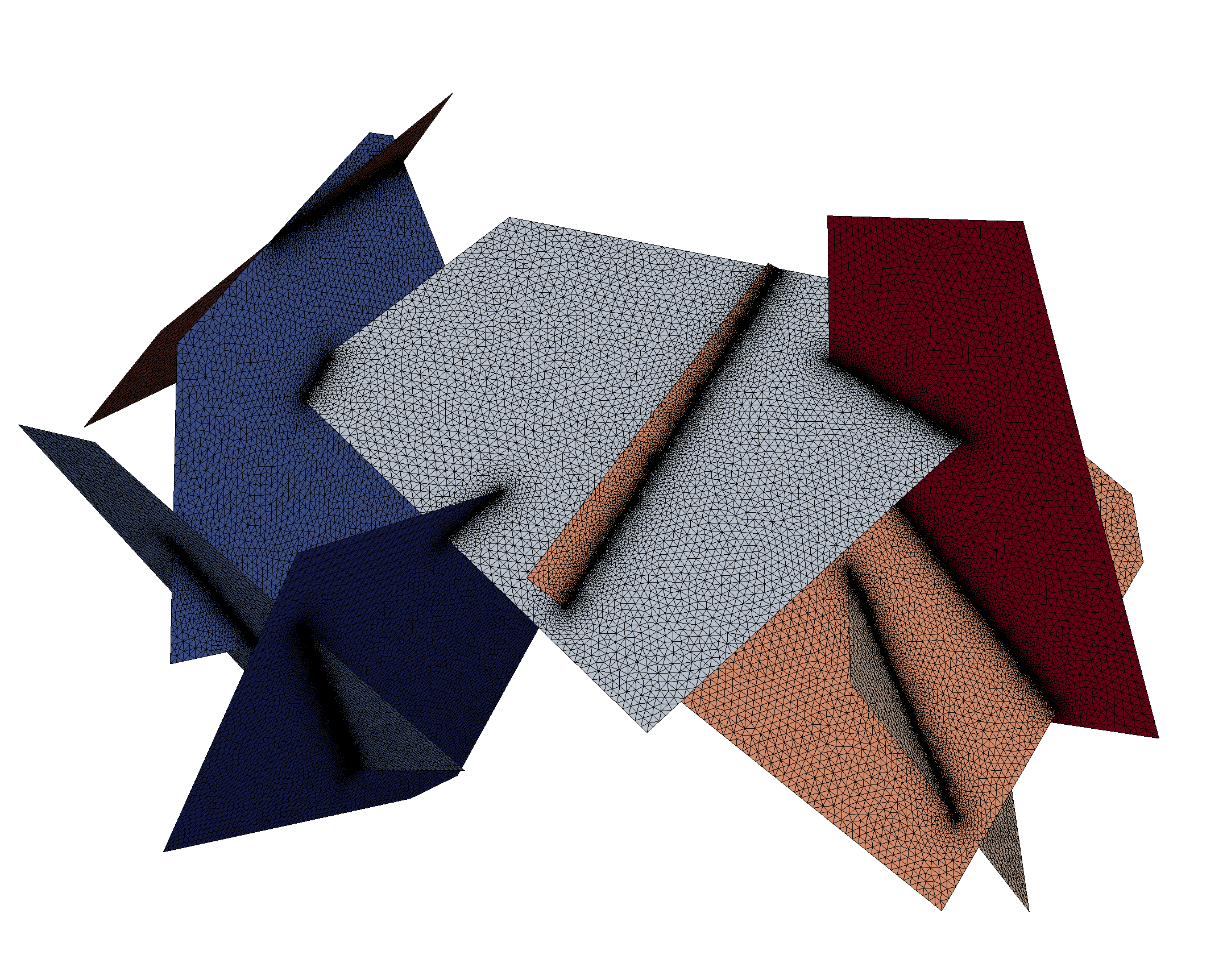}}
    {\includegraphics[width=0.475\textwidth]
      {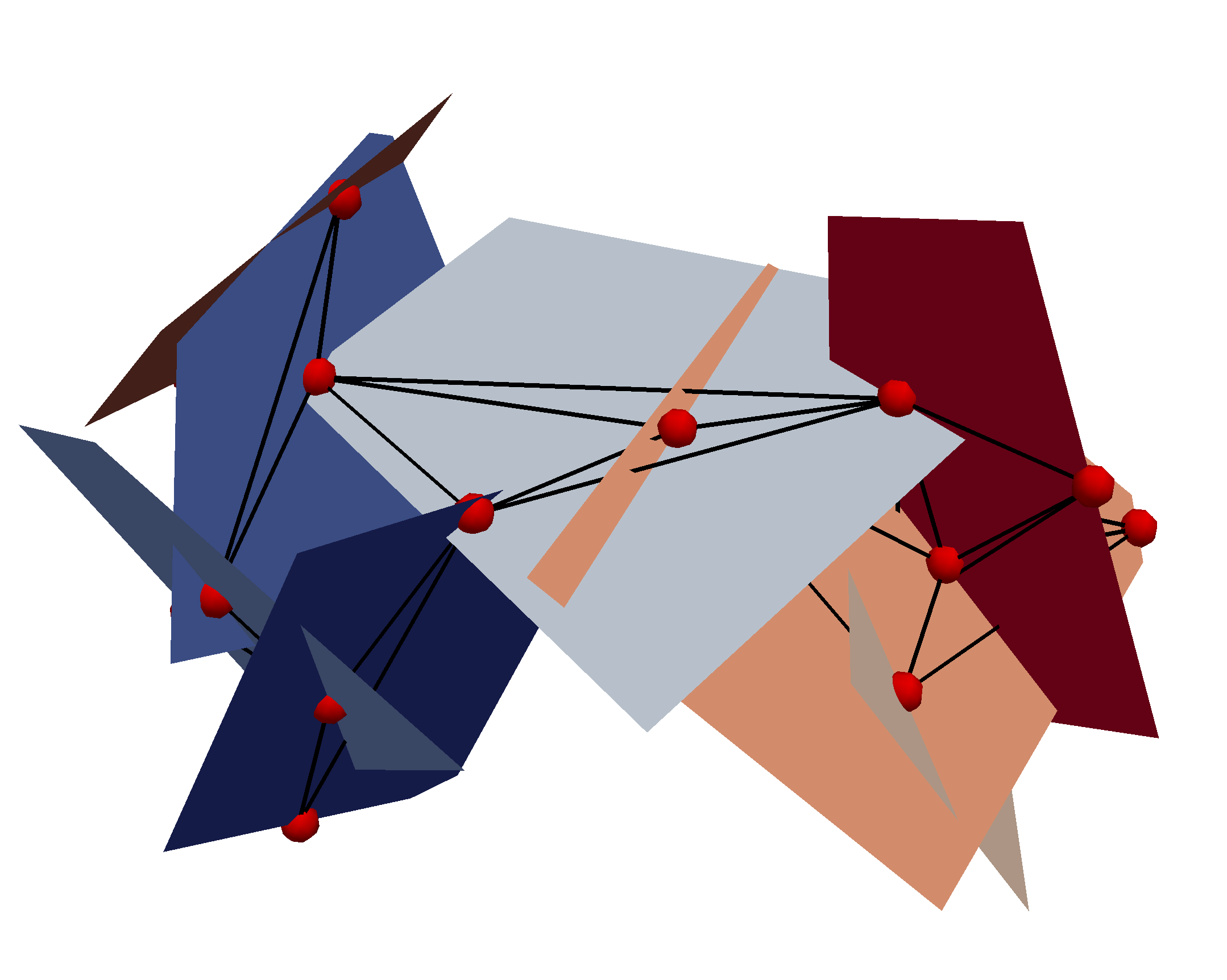}}
  \caption{The general 
  workflow in our proposed method
  involves building an equivalent graph for a given DFN. The connectivity of
  DFN is transformed into the graph connectivity. (Left) Eight fracture 
  DFN with a mesh that is used for performing the high-fidelity flow and
  transport calculations. (Right) Equivalent graph with nodes (red spheres) representing
  fracture intersections. The geometric information
  of the fractures such as distance between intersections, apertures, etc.,
  are stored in weights of the edges between the graph nodes. Properties
  such as permeability, porosity and viscosity are also stored in these 
  weights. The mesh to resolve the full network has 79792 triangular elements
  with 88200 vertices, while the graph representation has 15 nodes.}  
  \label{Fig:workflow}
\end{figure*}

In the last ten years there have been major advances in 
DFN simulation capabilities and high-fidelity simulations on 
large explicit three-dimensional fracture networks is now possible. 
One major challenge with the DFN approach that needed attention
is generating conforming meshes
that can resolve the small features resulting from the stochastic
creation of the networks. Methods such as the feature rejection algorithm for meshing (FRAM)
\citep{hyman2014conforming} have been proposed to overcome this issue 
effectively, which generates a mesh that is fine at an intersection and 
becomes increasingly coarse away from an intersection. 
Other research teams have opted to use mortar methods~\cite{pichot2012generalized}
or extended finite elements~\cite{berrone2013simulations} 
to alleviate the problem of having 
conforming meshes within fracture planes along intersections. 
The advantage of conforming meshes is that particle tracking
methods~\citep{makedonska2015particle}
can be used to simulate transport in a more natural 
way, which skirts the undesirable, yet common 
issues associated with numerical dispersion when resolving 
transport on unstructured meshes 
in an Eulerian framework. 

Even with these advances, the number of mesh cells
grows with the number of fractures that are included in the network. 
Even for a modest sized DFN with about 6300 
fractures, as shown in Fig.~\ref{Fig:dfn_example}, the number of unknowns
(degrees of freedom or dofs, hereafter) to solve flow are nearly 13 million. 
For target applications where the range of length scales can range up
to four orders of magnitude~\cite{bonnet2001scaling},
the number of dofs can be in the billions. 
A common workaround is to not include fractures below a given length scale. 
However, while ignoring these smaller-scale fractures gives reasonable first breakthrough predictions,
the tails tend to be inaccurate. For example, Karra et al.~\cite{karra2015effect} have shown
that for improving production curve tail estimates one needs to incorporate
smaller-scale fractures, that are typically ignored. Such large dof domains may be solvable using high-performance 
computing (HPC) software, for instance, using dfnWorks 
\citep{hyman2015dfnworks} for DFN
generation and PFLOTRAN \citep{lichtneretal2013PFLOTRAN} 
for solving flow and transport.
Even then, the stochastic nature of the models dominate the flow and transport
behavior that are only known in a statistical sense, and hence one has to
account for uncertainty. 
However, incorporating such large domains
in an uncertainty quantification (UQ) framework,
where hundreds (or more) of such realizations have to be run, is computationally 
intractable, not to mention, processing the copious amounts of data generated 
would be challenging.

We present a model-reduction technique to reduce the computational
complexity by solving flow and transport on a graph representation of a DFN. 
The topology of the nodes and edges of the graph is determined by the fracture network
and weights on nodes and edges seek to capture geometric 
and hydraulic properties of the fracture planes. 
We adopt a mapping where each intersection in the DFN is represented by a node on the
graph, which ensures that the connectivity of the DFN is maintained. The geometrical
information of the fractures such as distance between the intersections,
fracture apertures, as well as flow and transport properties, such as 
permeability and porosity, are incorporated in weights assigned
to the edges connecting the nodes. Additional nodes are placed in the graph
to incorporate boundary conditions at the inflow and outflow boundaries.
The idea behind solving
on an equivalent graph is that: (i) the number of dofs to be solved depend on the 
number of nodes on the graph which in our case will depend on the number of
fracture intersections, and (ii) we avoid meshing on each individual 
fracture which is a highly time-consuming step in a DFN model construction.
Now that high-fidelity flow and transport simulations on explicit 
three-dimensional DFN can be performed at large scales,
it provides us the opportunity to examine how the 
simplifying assumptions used in the low-order models influence the 
computational burden and quantities of interest. We use 
our in-house developed dfnWorks HPC suite for this purpose.
In particular, we aim to address the trade-off between computational speed and
accuracy relative to the fully resolved networks. Furthermore, by performing
accuracy studies, we can infer how much correction one
needs to make on the graph-based reduced model predictions.

It is worth noting that recent applications of graph theory
to fracture networks have helped gain insight into the structure
and connectivity of these networks. Valentini et al.~\cite{valentini2007small}
were one of the first ones to use graph equivalent
of natural fracture systems to study their features.
Andresen et al.~\cite{andresen2012topology} have mapped two-dimensional 
fracture outcrops from south-east Sweden
into graphs, and used various graph-based metrics such as 
clustering and efficiency to study their topology and connectivity. 
Santiago et al.~\citep{santiago2016descriptive} 
have developed an algorithm to process images of two-dimensional 
outcrops into graphs and used graph theory centrality measures to identify
key nodes for flow. Hyman et al.~\cite{hyman2017passage} used graph-based
techniques to identify subnetworks that give similar first passage time as
the full DFN. However, with their approach one needs to still solve 
flow and transport
on the DFN-equivalent of the subnetwork.
Ghaffari et al.~\cite{ghaffari2011fluid} have mapped 
two-dimensional fracture networks 
into graphs with fractures represented as nodes and their intersections 
being edges on the graphs, similar to Andersen et al.~\cite{andresen2012topology}. 
They then solved for steady flow on this graph by solving the graph Laplacian
to calculate the velocity distribution in the network. However, their work
was restricted to two-dimensional fracture networks while we focus on
more realistic three-dimensional fracture networks.
Furthermore, we are the first to compare 
the graph-based reduced model and the high-fidelity DFN model,
in terms of accuracy as well as computational performance.

We find that that solving flow and transport on 
the equivalent graph is $\mathcal{O}(10^4)$ times faster, thereby
one can feasibly incorporate a DFN model with a wide range of fracture sizes 
from millimeters to kilometers, within a
UQ framework. We show good accuracy for small networks while for 
larger networks where small-scale heterogeneity is more prominant, deviations 
from the high-fidelity DFN results are observed. For the larger networks,
we show that the graph-based approach generally over-predicts tracer
breakthrough times, always within an order of magnitude of the DFN predictions.
The systematic bias in the graph method, makes it amenable to UQ 
correction techniques.

In this paper, by flow we mean flow of a fluid
(e.g., water) in a fractured porous medium,
and by transport, we mean transport of a conservative tracer in this 
flow field. The paper is organized as follows. 
A brief overview of the DFN approach, 
the governing equations, and solution methodology
used to solve these governing equations on a given DFN, are detailed in 
Sec.~\ref{Sec:S2_DFN}. Details of the DFN to graph mapping methodology
along with the flow and transport solution algorithm on the equivalent
graph are discussed in Sec.~\ref{Sec:S3_Graph}. Breakthrough 
curves obtained using the full DFN and the equivalent graphs
are compared and analyzed in Sec.~\ref{Sec:S5_Comparison}.
Finally, conclusions are drawn in Sec.~\ref{Sec:S6_Conclusions}.
%
\section{Methodology}
\label{Sec:S2_Methodology}
In this section, we give an overview of the methods used to generate DFNs,
and to solve flow and transport on them. We also discuss the algorithm for
solving flow and transport on a graph along with the method we developed
to convert a DFN to an equivalent graph.

\subsection{Discrete Fracture Network}
\label{Sec:S2_DFN}
The computational suite dfnWorks \citep{hyman2015dfnworks} is used for
DFN generation, meshing, and solving flow and transport on DFN.
The approaches used to generate DFNs, and to solve flow and transport 
using dfnWorks are briefly described in this sub-section. For more
details, we refer the interested reader to \citep{hyman2015dfnworks}.

\subsubsection{Generation and Meshing}

Statistical distributions of fracture characteristics taken from field
measurements are
used to stochastically generate fractures. Characteristics include
size, location, aperture and orientation. Individual fractures are then meshed
using LaGriT toolkit \citep{lagrit2011}. Care is taken to ensure that 
the meshes are conforming at the intersections using the feature rejection algorithm (FRAM) \citep{hyman2014conforming}. 
FRAM uses a minimum length that is user defined for feature representation 
in the DFN. All the geometric features below the minimum length are not resolved. The 
algorithm also generates meshes that are fine at the fracture intersections
to resolve the smaller features for accuracy
and coarsens away from the intersections, thereby reducing the overall
number of grid cells and computational resources needed.

\subsubsection{Flow}
The generated and meshed DFN is then used to solve for steady state flow. The governing equation solved is a result of 
balance of mass and Darcy's model, given by \citep{bear2013dynamics}:

\begin{align}
\nabla \cdot \left( k(\boldsymbol{x}) \nabla p \right) = 0,
\label{eq:steady_flow}
\end{align}
where $k$ is the spatially varying permeability 
and $p$ is the liquid pressure. Equation~\eqref{eq:steady_flow}
is numerically integrated using a two-point flux finite volume method,
subject to pressure boundary conditions at the inlet and outlet boundaries.
We use the subsurface flow solver
PFLOTRAN \citep{lichtneretal2013PFLOTRAN} for this purpose. To get an accurate
solution that maintains local mass balance,
PFLOTRAN reads Voronoi control volumes for the DFN Delaunay triangular
mesh. Voronoi meshes, by construction, ensure that the line
joining two cell-centers is perpendicular to the face between the 
the two control volumes, leading to accurate two-point flux calculations. 
LaGriT is used to perform the conversion from Delauney to Voronoi.

\subsubsection{Transport}
\label{sec:dfn_transport}
The particle tracking approach 
is used to calculate the breakthrough curves
of a conservative tracer in the flow field governed by 
Eq.~\eqref{eq:steady_flow}. 
Trajectory $\boldsymbol{x}(t)$ of a given particle is evaluated by 
integrating the kinematic equation

\begin{align}
\label{eq:pathline}
\frac{d \boldsymbol{x}(t)}{d t} = \boldsymbol{v}\left(\boldsymbol{x}(t)\right),
\quad \boldsymbol{x}(0) = \boldsymbol{x}_{\mathrm{init}},
\end{align}
where $\boldsymbol{x}_{\mathrm{init}}$ is the initial position of the particle. The time
taken for the particle to travel from the inlet to the domain outlet, 
is then calculated.
For solving Eq.~\eqref{eq:pathline}, one needs a particle's velocity vector at 
every location, which is related to Darcy velocity vector $\boldsymbol{q}$
at that location via

\begin{align}
\label{eq:velocity}
\boldsymbol{v}\left(\boldsymbol{x}\right) = 
\dfrac{\boldsymbol{q}\left(\boldsymbol{x}\right)}{\varphi},
\end{align}
where $\varphi$ is the porosity, that can be assumed to be fairly constant in rock.
A uniform mass is assigned to each particle.

Since two-point flux finite volume formulation gives only the
normal component of the Darcy velocity
$q_n$ from the pressure solution at the Voronoi cell-centers via the Darcy model:

\begin{align}
q_n := \boldsymbol{q} \cdot \boldsymbol{n} = -k \left(\boldsymbol{x}\right)
\nabla p \cdot \boldsymbol{n},
\end{align}
where $\boldsymbol{n}$ is the unit normal, a velocity reconstruction method \citep{painter2012pathline} 
is used to calculate velocity vectors
at center of the Voronoi control volumes (which are vertices of the
corresponding Delaunay mesh). Once the Darcy velocity vector
$\boldsymbol{q}$ is known at the 
Delaunay vertices, Eqs.~\eqref{eq:pathline},~\eqref{eq:velocity} are used to
integrate for the particle pathlines. A predictor-corrector method is used
to perform this integration. Details of the particle tracking method used for
DFN can be found in \citep{makedonska2015particle}.

\begin{figure}
    \includegraphics[width=0.35\textwidth]
      {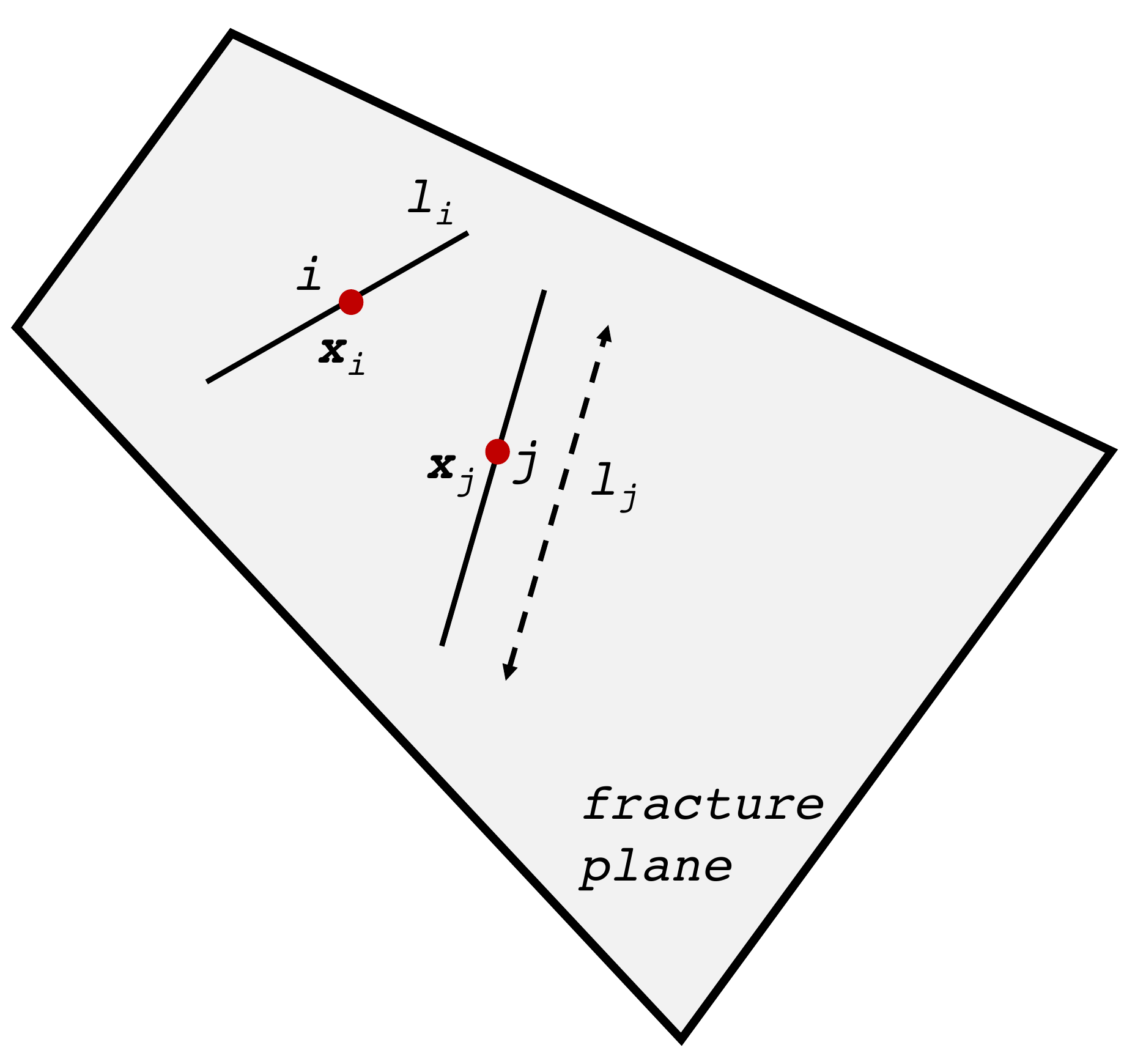}
  \caption{Illustration 
  of a single fracture
  plane showing how the geometrical information of fractures is used to
  map the intersections $i$, $j$ to nodes of an equivalent graph.}
  \label{Fig:cartoon_frac_plane}
\end{figure}
%

\subsection{Graph Flow and Transport Algorithm}
\label{Sec:S3_Graph}

In this sub-section, we present the mapping 
between DFN and graph that we adopt. 
Then we derive general flow governing equations on
a graph followed
by a description of the approach to solve these equations. 
The methodology used to
calculate the conservative tracer transport breakthrough on a graph
from the flow solution on the said graph is finally described.

\subsubsection{Discrete Fracture Network to Graph Mapping}
\label{Sec:mapping}



Consider a fracture plane with two intersections $i$ and $j$, such as those shown in 
Fig.~\eqref{Fig:cartoon_frac_plane}. We build a graph $G$ with nodes $i$, $j$
corresponding to these intersections while the edge on the graph corresponds
to the fracture plane. A node is added to the graph for each inflow or 
outflow plane.
Edge weights $w_{ij}$ on the graph are based on geometric and 
hydrological properties of the fracture plane.  
Figure~\ref{Fig:workflow} illustrates 
the workflow of converting a DFN into an equivalent graph
for an eight fracture network. 
Nodes are shown as red spheres and edges are black lines. 
The mesh to resolve the full network has 179792 triangular elements
with 88200 vertices, while the 
graph representation has only 15 nodes.

\subsubsection{Flow}
\label{Sec:graph_flow}
Let $N$ be the number of nodes in $G$.
Assuming steady flow, the balance of mass for
the fluid at a node $i$ in $G$, can be written as

\begin{align}
\label{eq:mass_graph}
\sum_{j=1}^{N} Q_{ij} = 0,
\end{align}
where $j$ is a node that is adjacent (or connected) to $i$, 
$Q_{ij}$ is the mass flux that flows through the connection $i$ to $j$. One 
can then relate $Q_{ij}$ to pressures $P_i$, $P_j$ at the nodes $i$, $j$, 
respectively, through an equivalent Darcy's model

\begin{align}
q_{ij} &= \dfrac{\kappa_{ij}}{\mu L_{ij}}\left(P_i - P_j\right), 
\label{eq:darcy_graph1}\\
Q_{ij} &= q_{ij}\alpha_{ij} \label{eq:darcy_graph2}, 
\end{align}
where $q_{ij}$ is the mass flux per unit area,
$\kappa_{ij}$ is the permeability of the fracture plane with intersections
$i$,  $j$ and $\mu$ is the viscosity.  
If $l_i$, $l_j$ be the lengths of the 
intersections, with $\boldsymbol{x}_i$, $\boldsymbol{x}_j$ being the 
centroids of the intersections (see Fig.~\eqref{Fig:cartoon_frac_plane}), 
and if $a_{ij}$ is the fracture aperture, 
then the area $\alpha_{ij}$ in Eq.~\eqref{eq:darcy_graph2} 
through which the fluid flows as it 
moves from $i$ to $j$ can be approximated to $a_{ij}\left(l_i + l_j\right)/2$.
Also, $L_{ij}$ in Eq.~\eqref{eq:darcy_graph1}
is set to the Euclidean distance between $\boldsymbol{x}_i$ and 
$\boldsymbol{x}_j$, $\|\boldsymbol{x}_i - \boldsymbol{x}_j \|$,
where $\| \cdot \|$ is the 
Euclidean norm. 

Equations~\eqref{eq:mass_graph},~\eqref{eq:darcy_graph1},~\eqref{eq:darcy_graph2}
imply that 

\begin{align}
\label{eq:mass_gov}
\sum_{j=1}^{N} w_{ij}\left(P_i - P_j\right) = 0,
\end{align}
where $w_{ij} := \dfrac{\kappa_{ij}\alpha_{ij}}{\mu L_{ij}}$.

Now, if we assign $w_{ij}$ as weights to edges of $G$, then one can define
an adjacency matrix \citep{newman2004analysis} $\boldsymbol{A}$  whose elements are $w_{ij}$. 
Note that when there is no connection between two nodes $p$ and $q$, 
the entry $A_{pq}$ is zero. Defining the degree of vertex
$m$ as $k_m := \sum_n A_{mn}$, one can 
re-write Eq.~\eqref{eq:mass_gov} conveniently, in the following matrix form

\begin{align}
\label{eq:mass_gov_matrix}
\left(\boldsymbol{D} - \boldsymbol{A}\right) \boldsymbol{P} = \boldsymbol{0},
\end{align}
where $\boldsymbol{D}$ is a diagonal matrix with elements $D_{mm} = k_m$,
$\boldsymbol{P}$ is a vector of pressure values $P_m$.

The matrix $\boldsymbol{L}:=\boldsymbol{D} - \boldsymbol{A}$ is the graph
Laplacian. In order to solve Eq.~\eqref{eq:mass_gov_matrix},
one needs to provide `boundary conditions' in terms of the pressure values 
at the inlet and outlet nodes. For a boundary node $b$, this is done by
setting $L_{bj} = \delta_{bj}$, where $\delta$ is the Kronecker delta, and
by replacing the $b$-th value in the $\boldsymbol{0}$ vector on the right hand
side of Eq.~\eqref{eq:mass_gov_matrix} with the known value of
the pressure at $b$. After solving for the pressure values
at the nodes, Eqs.~\eqref{eq:darcy_graph1},~\eqref{eq:darcy_graph2}
are used to evaluate the mass flux of water through the graph edges.

\subsubsection{Transport}
\label{Sec:graph_transport}
To calculate the breakthrough of a conservative tracer traveling from the inlet
to outlet nodes on $G$, we propose a method that is along the lines of the
particle tracking method.
The steps for this method are:
\begin{enumerate}
	\item The mass flux per unit area ($q_{ij}$)
	of water on the graph edges is first calculated.
	\item For a particle traveling from node $i$ to node $j$, the particle's
	velocity is then calculated as $v_{ij} = \dfrac{q_{ij}}{\varphi_{ij}}$,
	where $\varphi_{ij}$ is the porosity assigned to the edge connecting nodes
	$i$, $j$.
	\item Once $v_{ij}$ is known, the time taken for 
	a particle to travel from node $i$ to node $j$ is calculated, via 

	\begin{align}
		t_{ij} = \dfrac{L_{ij}}{v_{ij}} = \dfrac{L_{ij}\varphi_{ij}}{q_{ij}}. 
	\label{eq:graph_tt}
	\end{align}
	In Eq.~\eqref{eq:graph_tt}, we assume that a particle takes a straight line 
	path over the distance $L_{ij}$.
	\item When a node $i$ has multiple connected nodes, in order to decide which node the
	particle has to travel to, 
	a probability proportional to $q_{ij}$ is assigned to the particle.
\end{enumerate}
In our calculations, we set $\varphi_{ij}$ to a constant value of $\varphi$
that is same as the value used in high-fidelity DFN simulations.


\begin{figure}
  \subfigure{\includegraphics[width=0.5\textwidth]
      {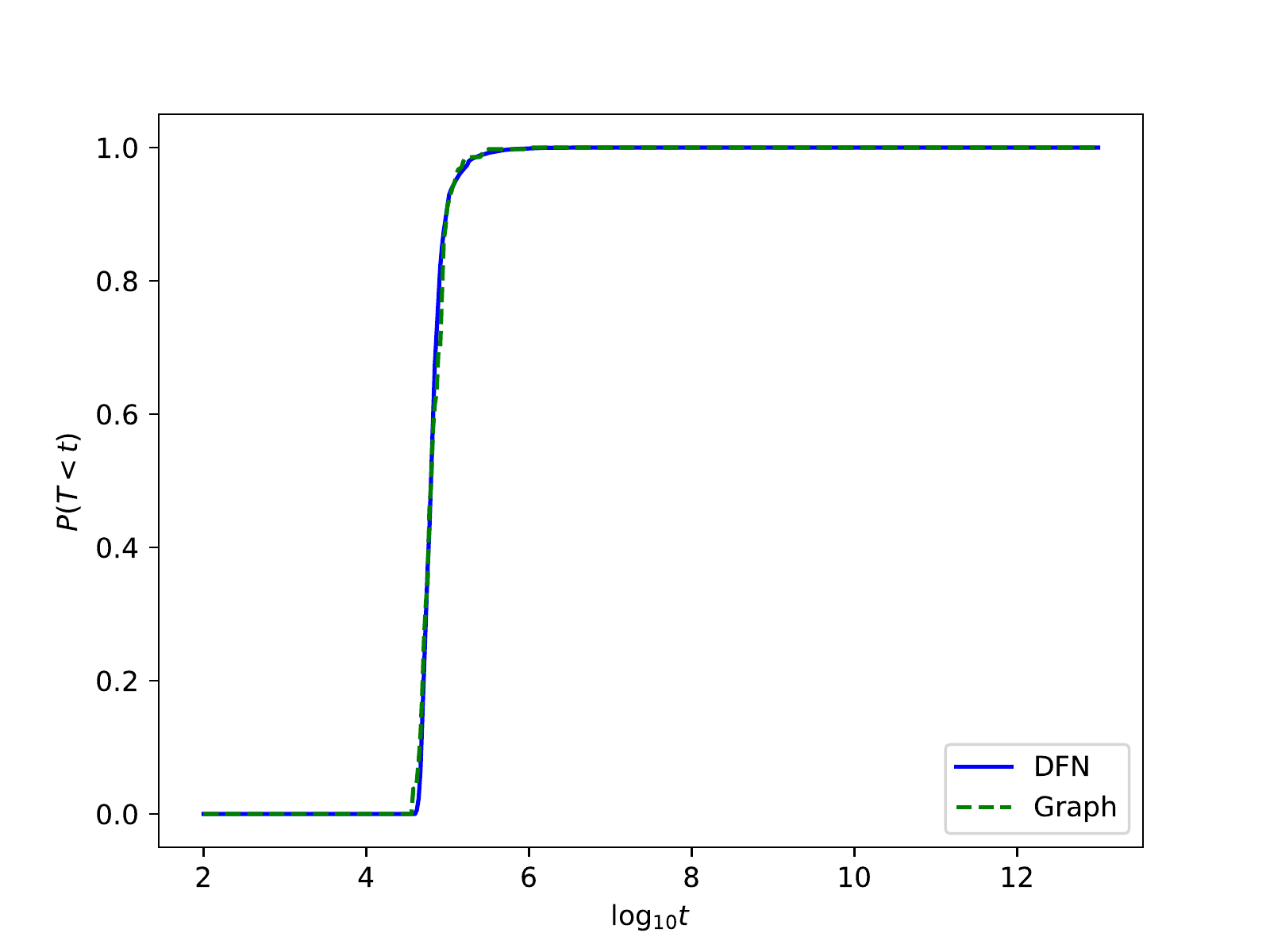}}
  \subfigure{\includegraphics[width=0.5\textwidth]
      {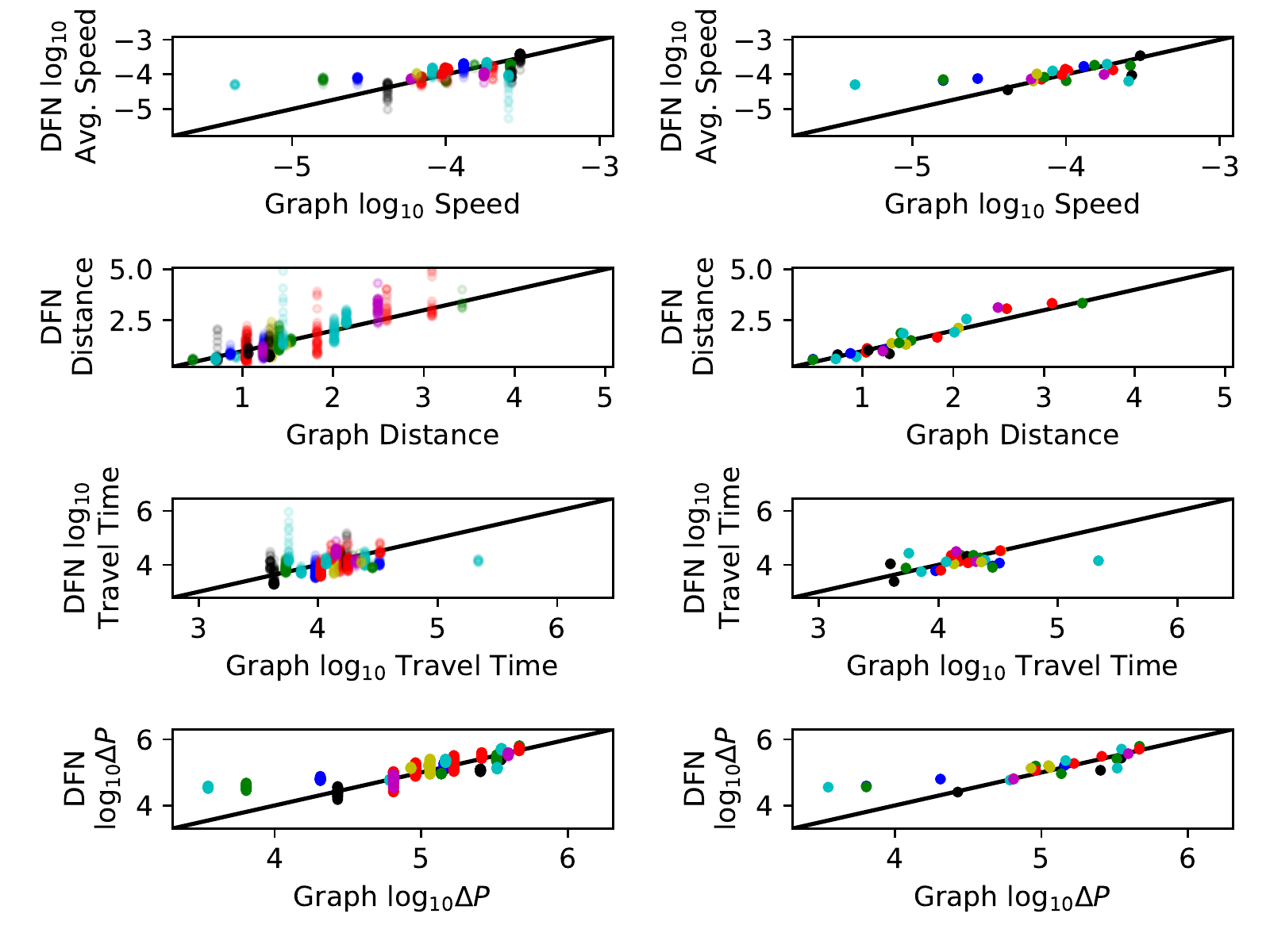}}
  \caption{Comparison between DFN and graph approaches for
  8 fractures with homogeneous permeability (Case 1). (Top) shows the 
  breakthrough curve comparison. Time is in seconds. (Bottom) shows the particle statistics between fracture intersections.
  The four subplots on the left side of (Bottom) are 
  individual particle statistics with all the particles traveling through
  the same connection shown with the same color. The four subplots on the 
  right side of (Bottom) are the average statistics of all the particles traveling
  through the same connection.}
  \label{Fig:8hom}
\end{figure}

\begin{figure}
  \subfigure{\includegraphics[width=0.5\textwidth]
      {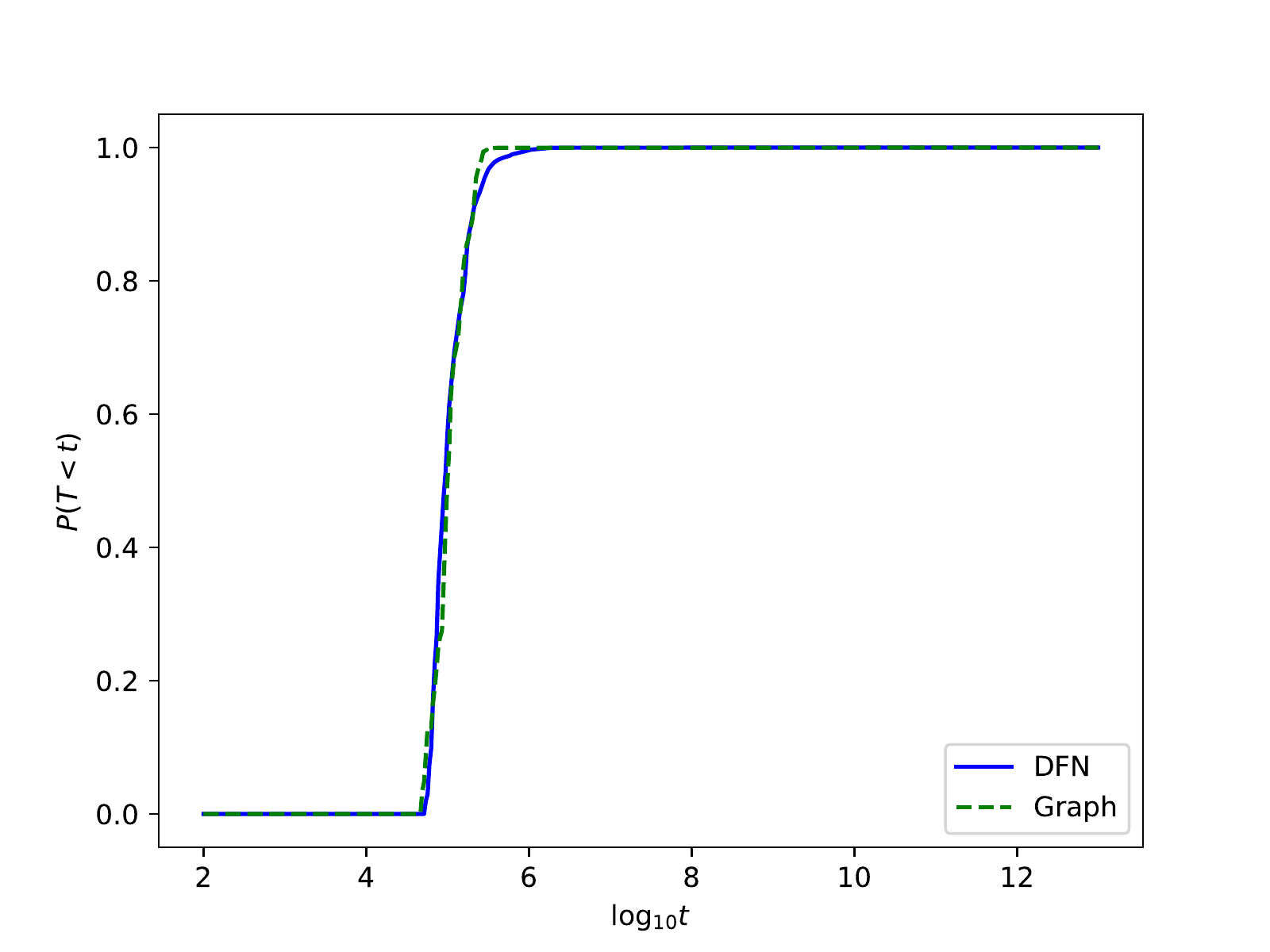}}
  \subfigure{\includegraphics[width=0.5\textwidth]
      {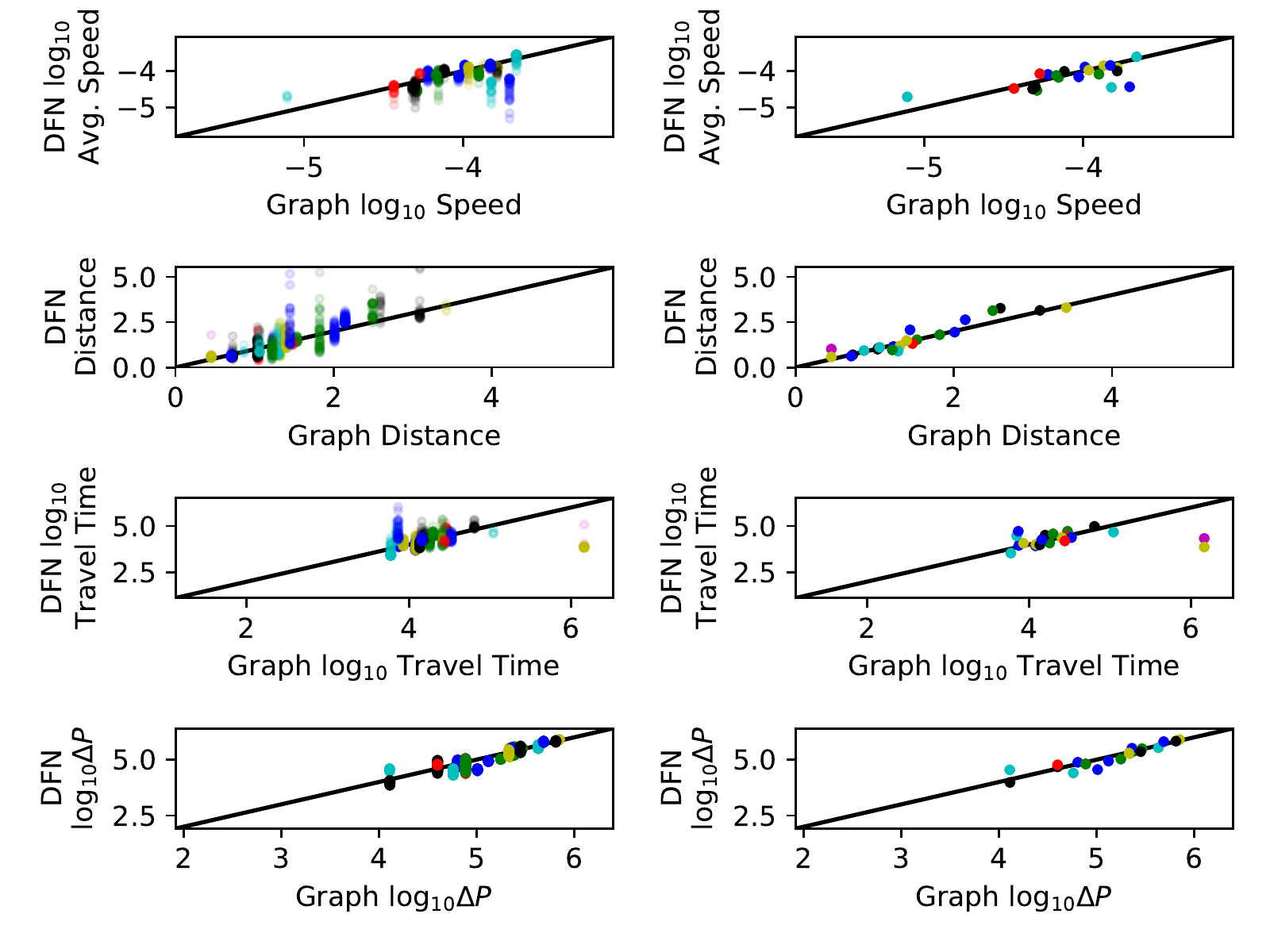}}
  \caption{Comparison between DFN and graph approaches for
  8 fractures with heterogeneous permeability (Case 2). (Top) shows the 
  breakthrough curve comparison. Time is in seconds. (Bottom) shows the particle statistics between fracture intersections.
  The four subplots on the left side of (Bottom) are 
  individual particle statistics with all the particles traveling through
  the same connection shown with the same color. The four subplots on the 
  right side of (Bottom) are the average statistics of all the particles traveling
  through the same connection.}
  \label{Fig:8het}
\end{figure}

\begin{figure}
  \begin{center}
  \subfigure{\includegraphics[width=0.5\textwidth]
      {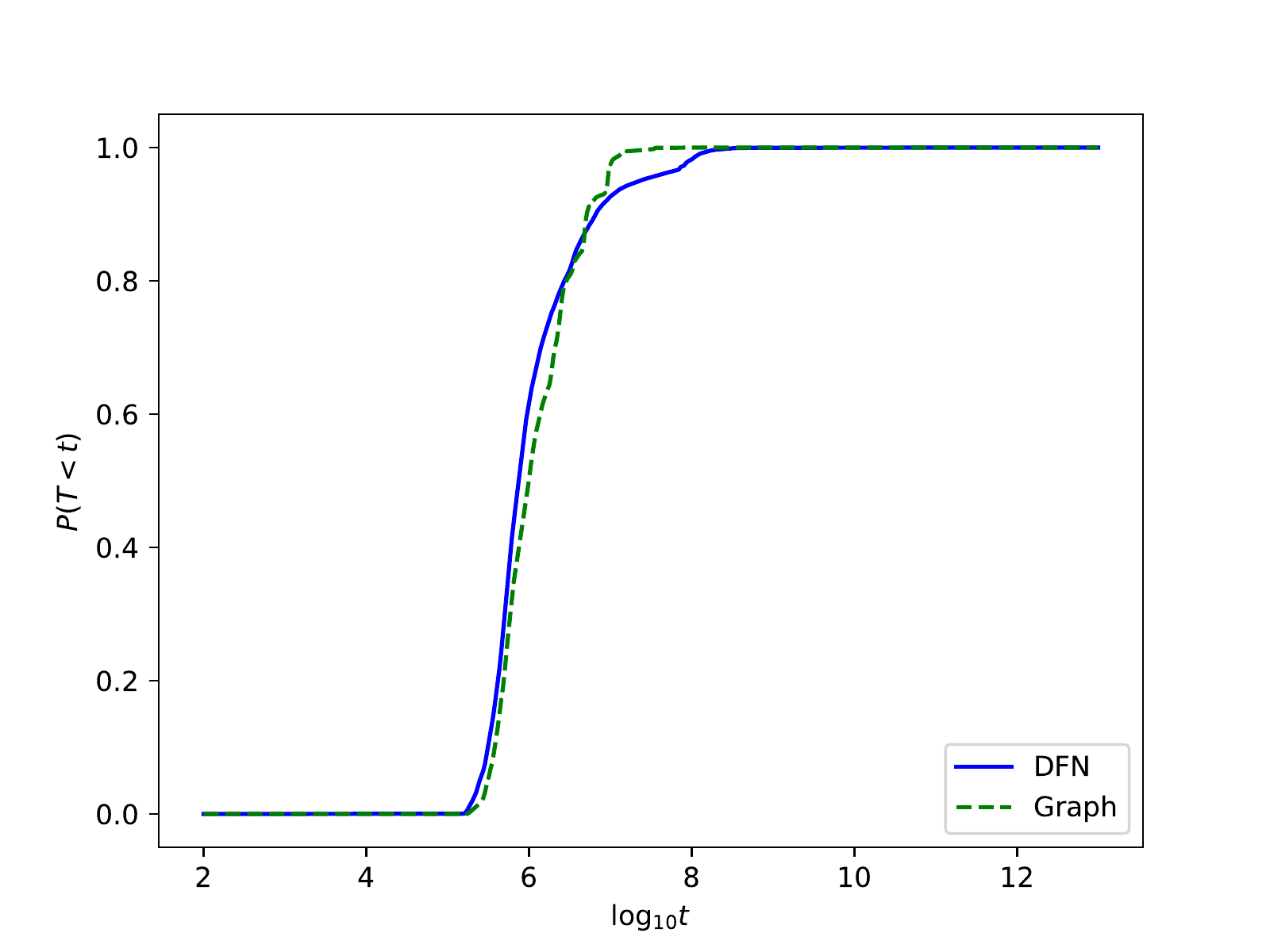}}
  \subfigure{\includegraphics[width=0.5\textwidth]
      {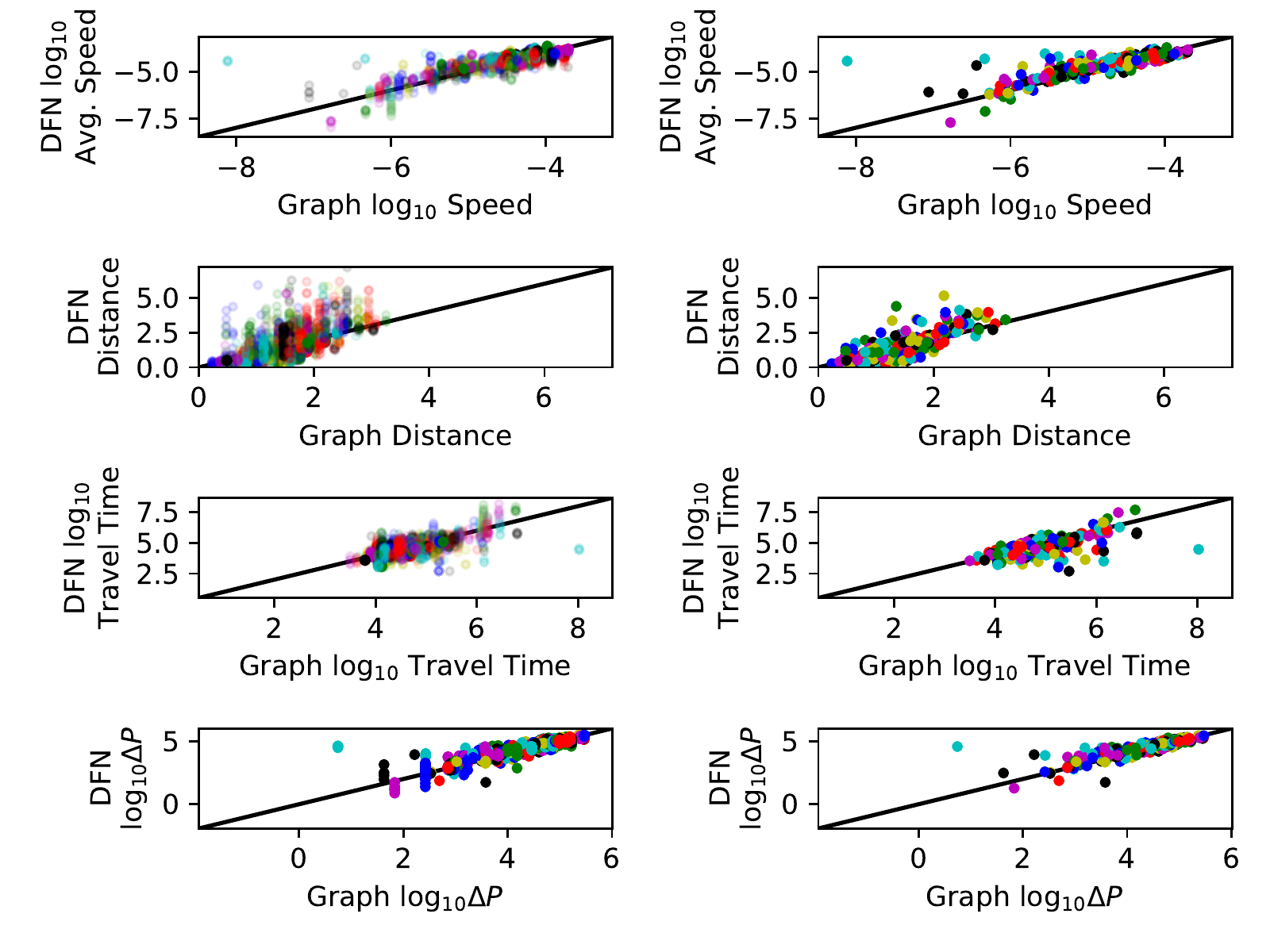}}
  \end{center}
  \caption{Comparison between DFN and graph approaches for
  150 fractures with homogeneous permeability (Case 3). (Top) shows the 
  breakthrough curve comparison. Time is in seconds. (Bottom) shows the particle statistics between fracture intersections.
  The four subplots on the left side of (Bottom) are 
  individual particle statistics with all the particles traveling through
  the same connection shown with the same color. The four subplots on the 
  right side of (Bottom) are the average statistics of all the particles traveling
  through the same connection.
  \label{Fig:150hom}}
\end{figure}

\begin{figure}
  \begin{center}
  \subfigure{\includegraphics[width=0.5\textwidth]
      {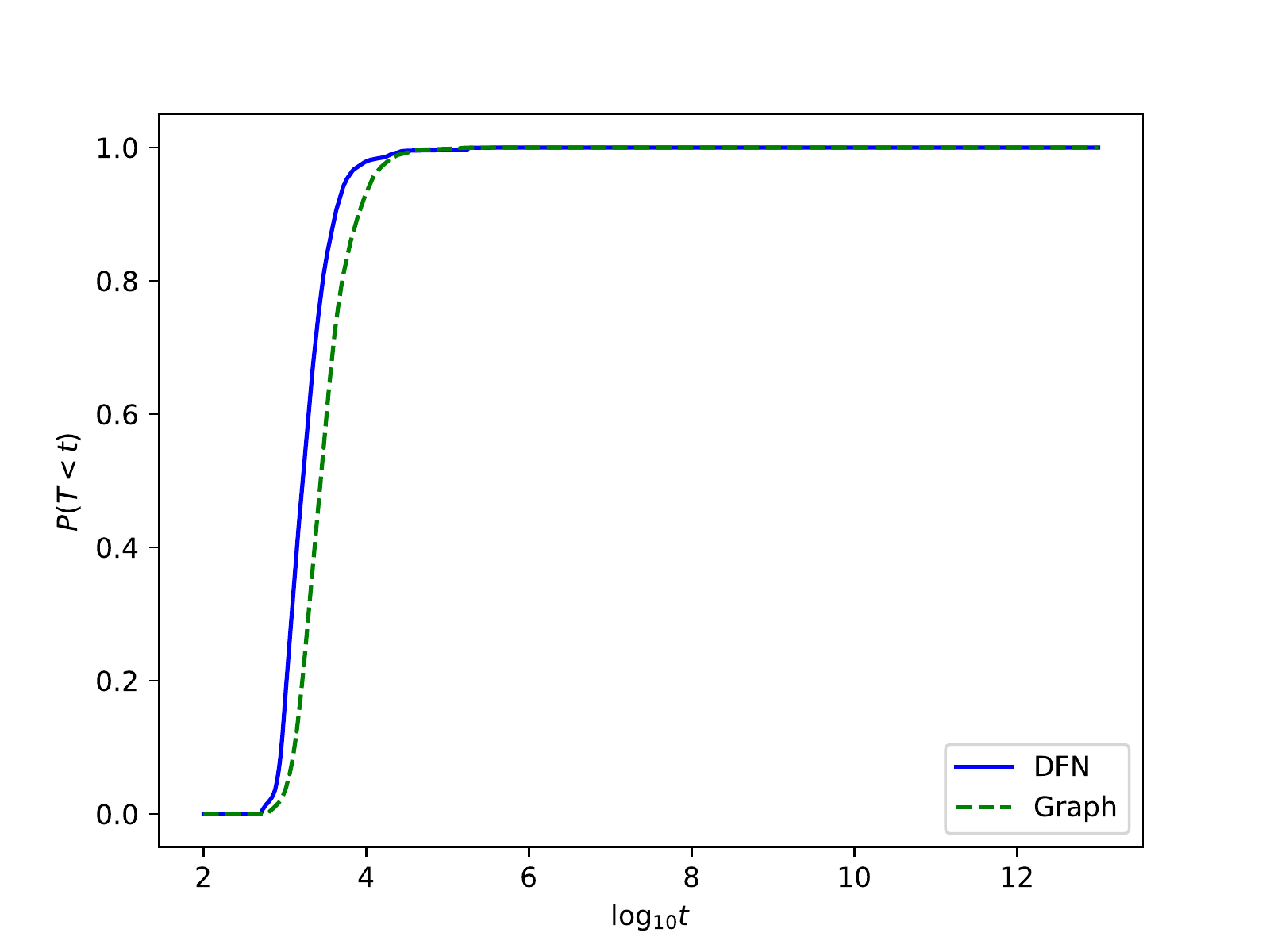}}
  \subfigure{\includegraphics[width=0.5\textwidth]
      {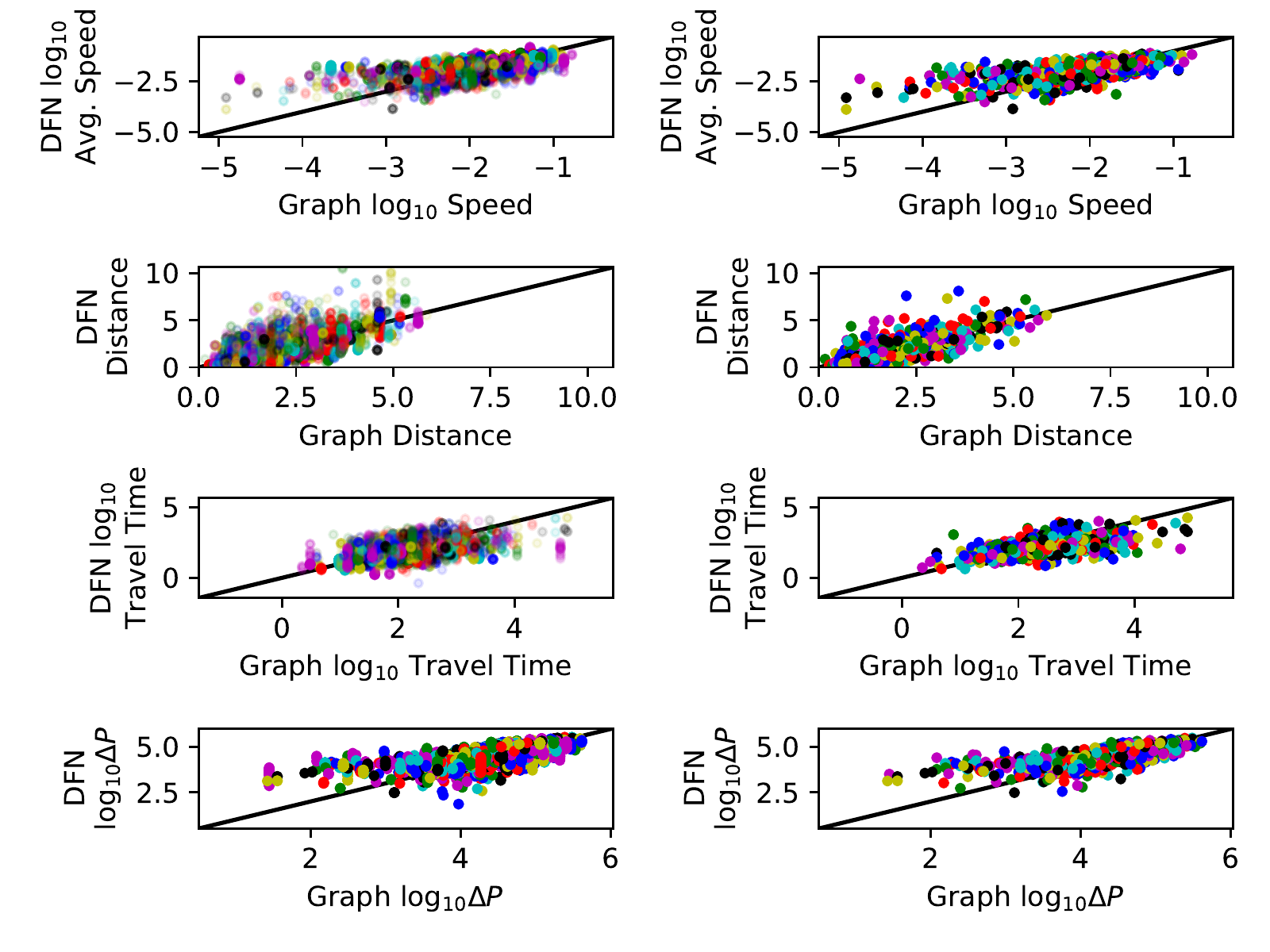}}
  \end{center}
  \caption{Comparison between DFN and graph approaches for
  500 fractures with heterogeneous permeability (Case 4). (Top) shows the 
  breakthrough curve comparison. Time is in seconds. (Bottom) shows the particle statistics between fracture intersections.
  The four subplots on the left side of (Bottom) are 
  individual particle statistics with all the particles traveling through
  the same connection shown with the same color. The four subplots on the 
  right side of (Bottom) are the average statistics of all the particles traveling
  through the same connection.
  \label{Fig:500het}}
\end{figure}

\begin{figure}
  \begin{center}
    \includegraphics[width=0.5\textwidth]
      {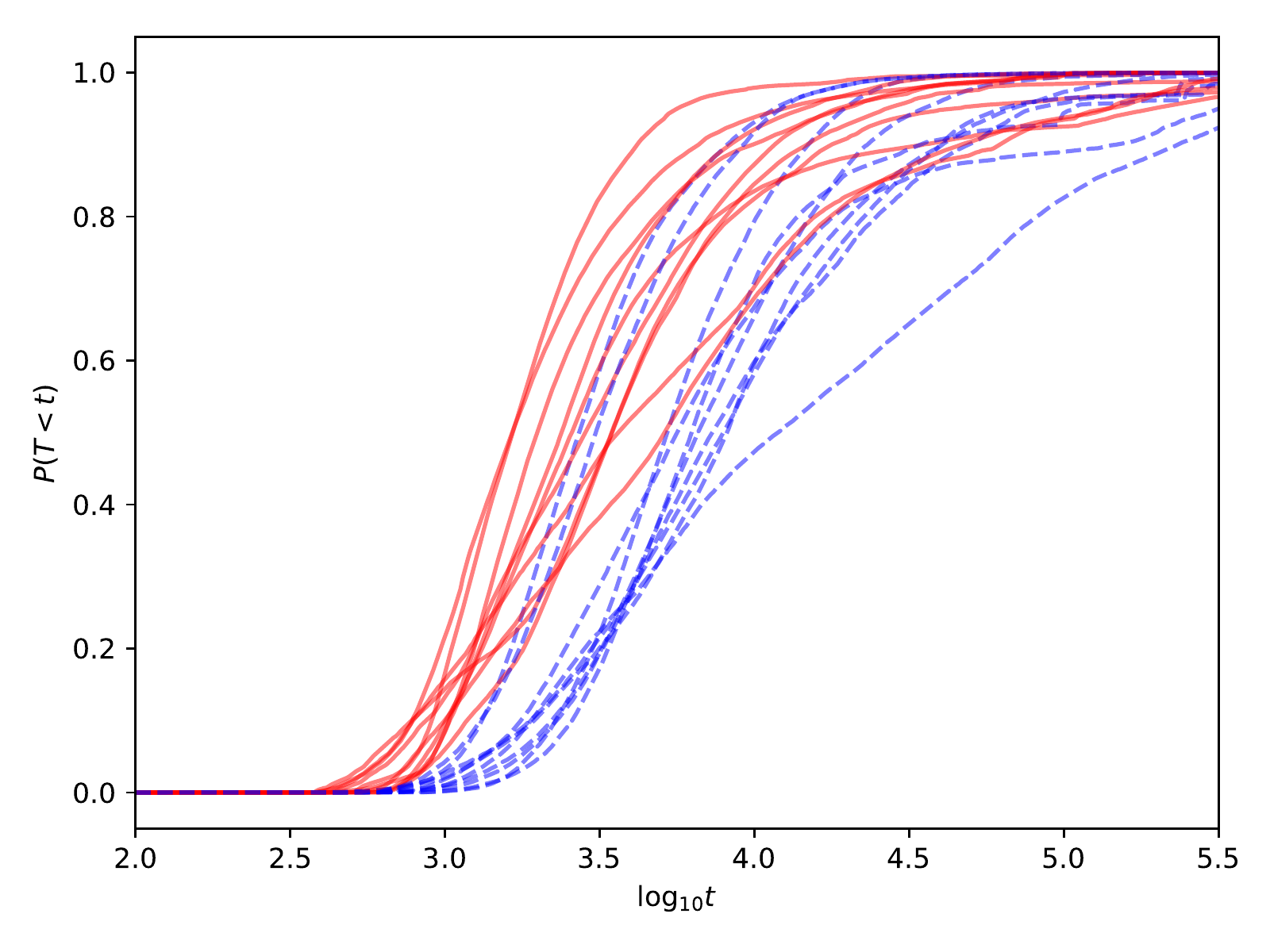}
  \end{center}
  \caption{Breakthrough curves for 
  10 realizations of 500 fracture networks with heterogeneous
  permeability. Blue curves are for graph and red is for DFN. The
  graph breakthrough is consistently slower than DFN. 
  \label{Fig:500het_10realiz}}
\end{figure}

\subsubsection{Transport bias correction}
For large networks, the breakthrough times predicted by the graph transport algorithm for particles tend to be biased in comparison to the DFN breakthrough times so that the breakthrough occurs later for the graph algorithm.
The bias will be discussed further in Sec. \ref{Sec:S5_Comparison} -- here we focus on how it can be corrected.
Simulating transport on these large networks is often computationally demanding, so it is important to note that our bias correction approach requires the use of a single high-fidelity DFN realization.
Other members of the ensemble from which the realization was drawn can then be accurately simulated using the graph model.

The basic approach to the bias correction is to use a power-law to improve the graph algorithm's prediction for the time to travel from one fracture intersection to another.
This is based on the ansatz that both the DFN and graph travel times follow a power law distribution \cite{berkowitz1997anomalous, berkowitz1998theory, metzler2000random}.
By examining a single high-fidelity DFN simulation in detail, we can obtain a wealth of information about the time to travel along a fracture from one intersection to another.
This is because particles typically travel through numerous fracture intersections and a DFN simulation tracks a large number of particles.
The power-law that is used takes the form
\begin{equation}
    t^c_{ij} = Ct_{ij}^\alpha
\end{equation}
where $t^c_{ij}$ is a corrected estimated of the time to travel from node $i$ to node $j$ in the graph and $t_{ij}$ is from Eq. \ref{eq:graph_tt}.
The power, $\alpha$ is estimated by a linear regression relating $\log t_{ij}$ to the corresponding values from the high-fidelity DFN realization.

%

\section{Comparison between DFN and Graph}
\label{Sec:S5_Comparison}
In this section, we compare breakthrough curves as well as CPU times between
the high-fidelity DFN runs and the graph approach.

\subsection{Breakthrough Comparison}
Breakthrough is a typical quantity of 
interest in subsurface flow and transport problems,
and hence we compare breakthrough curves and 
quantify the differences seen. For the purposes of this comparison, 
we construct four fracture networks with varying degrees of 
complexity. 
In all cases fracture centers are uniformly distributed throughout the 
domain and orientations are also uniformly random. 
The four cases with corresponding breakthrough comparison plots
are: 
\begin{itemize}
  \item \textit{Case 1:} 8 uniformly sized square fractures (side 
  length 3 meters) with permeability 
  being the same on all the fractures   (Fig.~\ref{Fig:8hom});
  \item \textit{Case 2:} The same network as in Case 1, but 
 with permeability varying between fractures. Permeabilities are sampled 
 from a log normal distribution with log variance of one, a moderate level
 of hydraulic heterogeneity. (Fig.~\ref{Fig:8het});
  \item \textit{Case 3:} 150 uniformly sized square fractures (side length of 1.5 meters)
  with same permeability on all fractures.
  (Fig.~\ref{Fig:150hom});
  \item \textit{Case 4:} Moderate sized network composed of approximately 500 circular fractures.
Fracture radii are sampled from a truncated power-law distribution with exponent $\alpha = 2.6$ and upper and lower cutoffs of 1 meter and 5 meters. 
The average $P_{32}$ value, total surface area over domain volume, of the networks is 2.78, a moderate network density. 
The permeability of the fractures is positively correlated to the fracture radius via a power-law relationship~\cite{hyman2016fracture}. (Fig.~\ref{Fig:500het});
\end{itemize}
Table~\ref{Tab:params} shows the parameters used in the flow simulations of the
four cases. To analyze the reason for any differences seen
between the two approaches, 
we have also plotted the statistics of flow and transport quantities
of individual as well as average of
particles traveling through each connection 
in Figs.~\ref{Fig:8hom}--\ref{Fig:500het}. 
Each connection here is the
connection between two intersections on a fracture, as described in 
Sec.~\ref{Sec:mapping}. These quantities include
distance traveled by a particle between any two intersections on fracture, 
the particle's speed as well
as the travel time over the distance, and
the pressure gradient across the two intersections.

The breakthrough curves match very well for both 
Case 1 (Fig.~\ref{Fig:8hom}) and Case 2 (Fig.~\ref{Fig:8het}), along 
with excellent correlation between the average DFN and graph particle 
flow and transport quantities. For Case 3, the graph predicts slower breakthrough
than DFN for the most part. The reason being graph under-predicts the 
pressure gradients across intersections by several orders of magnitudes 
(note the log scale in pressure gradient data),
and thus the particles traveling on these connections 
have several orders of magnitude slower
speeds and longer travel times. However, towards the end, DFN breakthrough is
slower. This is because there are some connections in the DFN where
the particles have to travel more distance, on an average, than the graph 
approach.
One possible reason for this is that DFN captures the pathline distances
of the particles while graph uses the straight line distance
between two fracture intersection centers, 
and so in some cases the average of the DFN pathline distances
between intersections is larger than the graph distance. In Case 4,
the graph consistently shows slower breakthrough due to several orders of
slower particle speeds and their travel times, similar to Case 3, 
but at a larger number of connections than Case 3. To check for 
consistency in the breakthrough comparison, we ran 10 realizations of 
Case 4. Fig.~\ref{Fig:500het_10realiz} shows the corresponding breakthroughs
with the graph being consistently slower than DFN. It is also seen that
as the number of fractures
increase, the under-prediction of the pressure gradients across intersections
increases with the graph based method
and thus the particles exhibit longer travel times.

\begin{table*}[]
\caption{\label{Tab:params}
Parameters used in both DFN and graph simulations.} 
\begin{ruledtabular}
\begin{tabular}{lcccc}
Quantity         & Case 1 & Case 2 &  Case 3 &  Case 4 \\ \hline
Number of connections &  15    &  15    &    216     &    575     \\
Inlet pressure   & 2 MPa &  2 MPa &  2 MPa &  2 MPa  \\
Outlet pressure  & 1 MPa &  1 MPa &  1 MPa &  1 MPa  \\
Log$_{10}$ Permeability     &   -12  &   [ -12.40, -11.60]        &   -12      &     [ -9.04, -9.68]   \\ 
No. of particles (graph) & 25,000    &  25,000    &  25,000   &  25,000   \\
No. of particles (DFN) & 25,000    &  25,000    &  25,000   &  25,000   \\
\end{tabular}
\end{ruledtabular}
\end{table*}

\begin{figure}
  \begin{center}
    \includegraphics[width=0.5\textwidth]{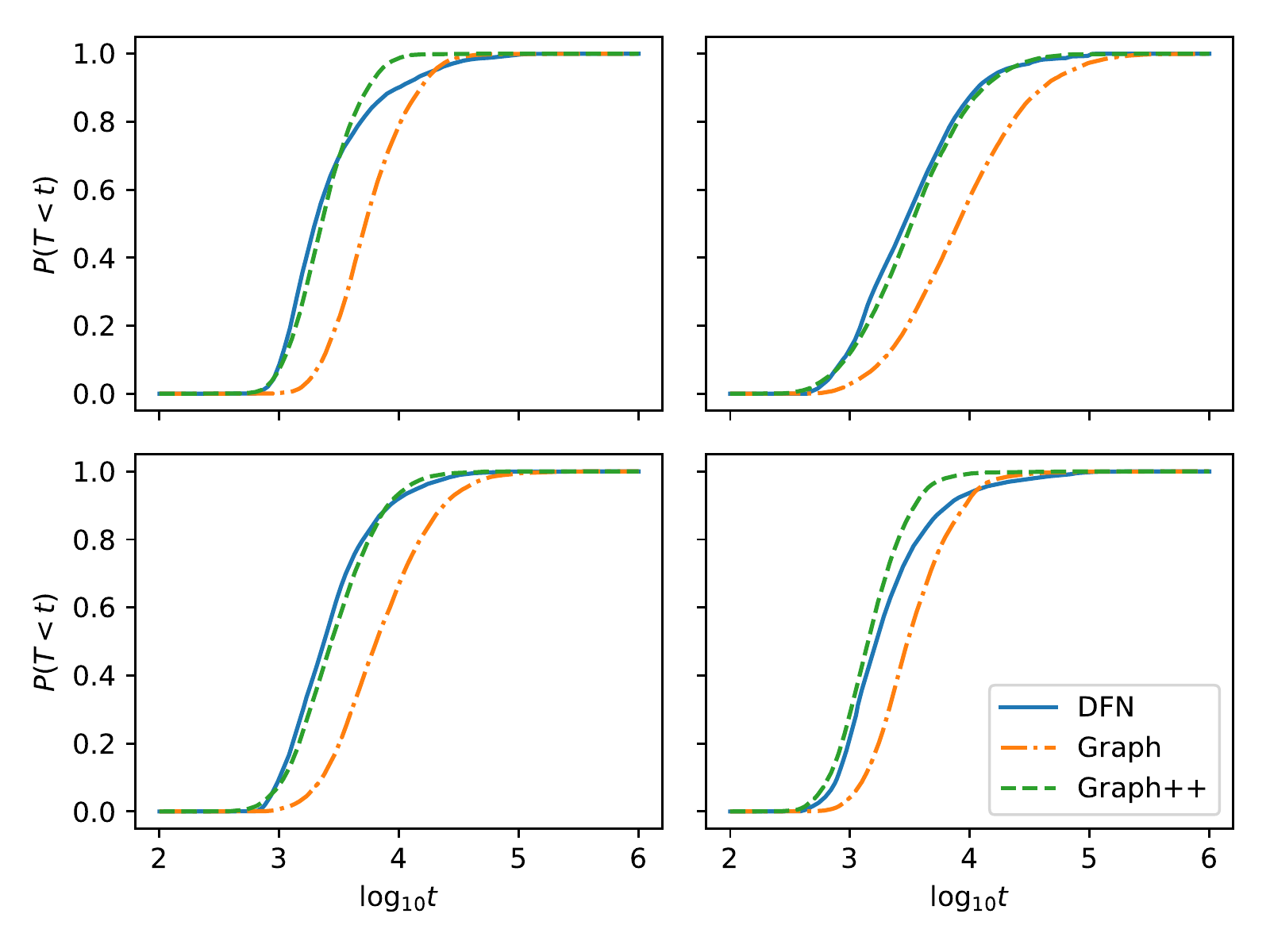}
  \end{center}
  \caption{Breakthrough curves for four realizations of 500 fracture networks with heterogeneous permeability. Blue curves are for the DFN, orange is for the graph, and green is the graph utilizing the bias correction procedure (called ``Graph++'' in the legend).}
\label{Fig:500het_correction}
\end{figure}

Using the bias correction procedure described previously, the accuracy of the predictions for Case 4 can be substantially improved.
Figure~\ref{Fig:500het_correction} shows the breakthrough curves for four realizations from the ensemble using the DFN, graph, and graph with bias correction (``Graph++'') models.
From this figure, it can be visually seen that the bias correction procedure significantly improves the accuracy of the graph model.
To quantify the improvement, we utilized the Kolmogorov-Smirnov statistic which is equal to the supremum of the difference between two cumulative distribution functions.
The expected Kolmogorov-Smirnov statistic for the graph model with the bias correction in comparison to the DFN model was approximately $0.09$.
Without the bias correction procedure, the expected Kolmogorov-Smirnov statistic was approximately $0.34$.
The bias correction procedure improves the Kolmogorov-Smirnov statistic and visually improves the fit.
From examining the trajectories, the largest errors tend to occur at later times (e.g., as can be seen in the upper left and lower right panels in Fig. \ref{Fig:500het_correction}), and is more accurate at earlier times.

\subsection{Computational Comparison}
\begin{figure}
  \begin{center}
    \includegraphics[width=0.48\textwidth]
      {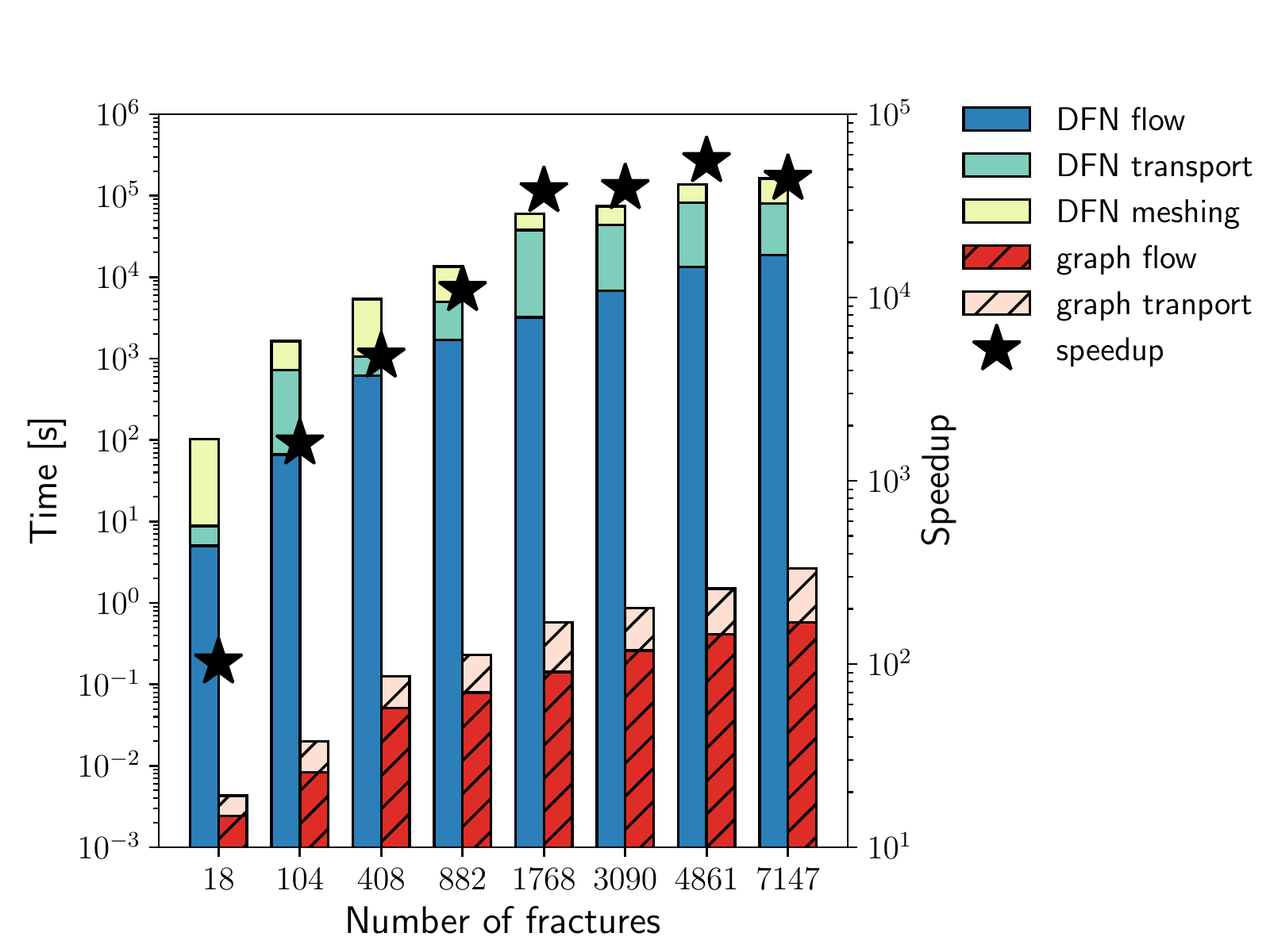}
  \end{center}
  \caption{Plot comparing the CPU times for various steps in  
  the graph and DFN methods.
  Note that the y-axis is in logarithmic scale. The star marker shows the
  ratio of graph method to DFN times.}
  \label{Fig:times}
\end{figure}

For comparing the computational performance of the graph-based and DFN 
approaches, networks with fractures increases from 18 to 7147, were used.
The CPU times for both the approaches with breakdown among the various
steps -- DFN meshing, flow and transport solve, 
graph flow and transport solve -- are shown for these networks
in Table~\ref{table:times}. 
Figure~\ref{Fig:times} shows these times as
histograms for one-to-one comparison along with the ratio
between total DFN time and total graph time shown as speedup. 
Networks for this comparison are composed of square fractures. 
The density of the networks is held constant and the size of the domain increased to increase the number
of fractures. 
All CPU times reported here were run with 1 
processor on a 32 core, 2 thread per core, AMD Opteron(TM) Processor 6272
with 528 GB RAM. Since the same DFN generation step is required
for both approaches, the CPU time for this step is not used in the comparison. 
The overall CPU times for the graph approach is up to $\mathcal{O}(10^4)$ times
smaller than DFN. The significantly 
faster times with the graph approach is due to
two factors: 1) meshing is the biggest bottleneck with 
the DFN approach and the graph approach avoids this step; 2) graph flow and transport
solves are at least three orders of magnitude faster than DFN due to 
significant ($\mathcal{O}(10^3)-\mathcal{O}(10^4)$) dof reduction.

\begin{table*}[]
\centering
\caption{CPU times on a single core for various steps in the DFN and graph approaches (shown in seconds).}
\label{table:times}
\begin{ruledtabular}
\begin{tabular}{llllllllll}
\multirow{2}{*}{No. of fractures} & \multirow{2}{*}{No. of cells} & \multirow{2}{*}{No. of trajectories} & \multicolumn{4}{c}{DFN} & \multicolumn{2}{c}{Graph}  \\
\cline{4-7} \cline{8-9}
& & & Generation & Meshing & Flow & Transport & Flow  & Transport \\
\hline
18  & 27415 & 498 & 0.03 & 92.52 & 1.01 & 5.02 & 0.002 & 0.002 \\
104 & 193308 & 1795 & 0.09 & 899.40 & 9.34 & 66.21 & 0.008 & 0.012 \\
408 & 780276 & 5891 & 0.43 & 4252.84 & 38.12 & 617.86 & 0.050 & 0.074 \\
882 & 1745002 & 8697 & 1.00 & 8451.90 & 95.41 & 1699.99 & 0.080 & 0.151 \\
1768 & 3581117 & 13724 & 1.57 & 22009.47 & 153.07 & 3210.52 & 0.142 & 0.439 \\
3090 & 6387657 & 19598 & 3.00 & 29931.83 & 260.21 & 6813.58 & 0.260 & 0.606 \\
4861 & 10232106 & 25988 & 5.85 & 55762.68 & 409.37 & 13269.95 & 0.410 & 1.080 \\
7147 & 15178277 & 41975 & 8.83 & 81392.85 & 592.63 & 18614.50 & 0.580 & 2.075 
\end{tabular}
\end{ruledtabular}
\end{table*}

%

\section{Conclusions}
\label{Sec:S6_Conclusions}
We successfully demonstrated that solving flow and transport on a graph equivalent to
a given DFN is 
$\mathcal{O}(10^4)$ times faster for large networks.
The graph approach takes advantage of the fact that: 1) each intersection
of a DFN is represented by a node and so the dofs are 
significantly smaller over DFN, and 2) meshing in fractures
is a time-consuming step in DFN and no meshing is needed in the graph approach.
Using breakthrough as the quantity of comparison, we 
compared the two approaches for various fracture networks with increasing
number of fractures. We found that graph approach reasonably predicts the breakthrough curves compared to DFN for smaller networks (8 fractures) and gives slower breakthrough for larger and more realistic networks with 150 and 500 fractures, with the graph prediction being
no more than an order of magnitude slower than DFN. We
found that this discrepancy is generally due to graph under-predicting
the pressure gradients across intersections on a fracture,
which leads to slower particle speeds between the intersections and longer
travel times. Furthermore, the systematic bias 
in the graph method over DFN, allows for performing corrections to 
the graph predictions.
We also developed a correction methodology to reduce the systematic bias, and showed that this methodology significantly improves the graph algorithm and gives results that are close to the high-fidelity DFN predictions. Overall, the speed of the graph approach along with the good accuracy using the proposed bias correction methodology, makes the graph
approach a promising model reduction technique for flow and transport in
fractured media. 
\begin{acknowledgments}
The authors thank Los Alamos National Laboratory LDRD program for the support
through project 20170103DR. The authors thank Nataliia Makedonska for help with 
DFN particle tracking code output needed for this paper.
\end{acknowledgments}

\bibliography{abbreviated}

\end{document}